\documentclass[conference]{IEEEtran}
\IEEEoverridecommandlockouts
\usepackage[noadjust]{cite}
\usepackage{amsmath,amssymb,amsfonts}
\usepackage{algpseudocode}

\usepackage[ruled,vlined,linesnumbered]{algorithm2e}    
\usepackage{graphicx}
\usepackage{textcomp}
\usepackage{xcolor}
\usepackage{mathtools}
\usepackage{amsmath}
\usepackage{caption}

\newtheorem{theorem}{Theorem}[section]

\newtheorem{lemma}[theorem]{Lemma}

\def\BibTeX{{\rm B\kern-.05em{\sc i\kern-.025em b}\kern-.08em
		T\kern-.1667em\lower.7ex\hbox{E}\kern-.125emX}}

\SetKwInput{KwData}{Input}
\SetKwInput{KwResult}{Output}
\SetKwFunction{FLoop}{convTrans}
\SetKwFunction{FLoopb}{conv1bTrans}
\SetKwProg{Fn}{Function}{}{}

\begin{document}
	
	\title{Simplified Successive Cancellation List Decoding of PAC Codes\\
	}
	
	\author{\IEEEauthorblockN{Hamid Saber}
		\IEEEauthorblockA{\textit{Samsung Semiconductor Inc} \\
			\textit{San Diego, USA}\\
			hamid.saber@samsung.com}
		\and
		\IEEEauthorblockN{Homayoon Hatami}
		\IEEEauthorblockA{\textit{Samsung Semiconductor Inc} \\
			\textit{San Diego, USA}\\
			h.hatami@samsung.com}
		\and
		\IEEEauthorblockN{Jung Hyun Bae}
		\IEEEauthorblockA{\textit{Samsung Semiconductor Inc} \\
			\textit{San Diego, USA}\\
			jung.b@samsung.com}
	}
	
	\maketitle
	
\begin{abstract}
Polar codes are the first class of structured channel codes that achieve the symmetric capacity of binary channels with efficient encoding and decoding. In 2019, Arikan proposed a new polar coding scheme referred to as \emph{polarization-adjusted convolutional (PAC)} codes. In contrast to polar codes, PAC codes precode the information word using a  convolutional  code prior to polar encoding. This results in material coding gain over polar code under Fano sequential decoding as well as successive cancellation list (SCL) decoding. Given the advantages of SCL decoding over Fano decoding in certain scenarios such as low-SNR regime or where a constraint on the worst case decoding latency exists, in this paper, we focus on SCL decoding and present a simplified SCL (SSCL) decoding algorithm for PAC codes. SSCL decoding of PAC codes reduces the decoding latency by identifying special nodes in the decoding tree and processing them at the intermediate stages of the graph. Our simulation results show that the performance of PAC codes under SSCL decoding is almost similar to the SCL decoding while having lower decoding latency. 
		
\end{abstract}
	
	
\section{Introduction}
	
Proposed by Arikan~\cite{PolarArikan}, polar codes are the first class of structured channel codes which achieve the capacity of binary input memory-less output symmetric channels under successive cancellation (SC) decoding. In spite of their capacity achieving property, the performance of polar codes at finite length under SC decoding is quite poor mainly due to weakness of SC decoding. Several improved decoders have been proposed to improve the finite length performance of polar codes. Among these decoders successive cancellation list (SCL) decoder shows the most promising performance~\cite{sclTV}. The performance was further improved by concatenating an outer CRC code to the polar code and perform a CRC-aided SCL (CA-SCL) decoding~\cite{CASCLKaiNiu}. Polar codes under CA-SCL decoder have been the state of the art polar coding scheme. 
	
In his Shannon lecture at at ISIT in 2019, Arikan proposed a new polar coding scheme, referred to as polarization-adjusted convolutional (PAC) codes. With PAC codes, an information word is encoded with a convolutional code (CC) prior to applying the polarization transform~\cite{ArikanPAC}. This simple modification appears to boost the performance of polar codes when used with Fano's sequential decoding. Similar performance can be achieved under SCL decoding with larger list sizes~\cite{pacSCLArman}.

In spite of its promising performance, Fano decoding has certain disadvantages, among which are the SNR-dependent time complexity with high complexity at low SNRs. This can be an issue at low-SNR regime where a worst case decoding latency needs to be ensured. In~\cite{pacSCLArman} and~\cite{pacSCLRowshan}, a SCL decoding was prospered for PAC codes based on the SCL decoding of polar codes with additional operations considering the CC precoder. SCL decoding of PAC codes demonstrates similar performance to Fano decoding at large list sizes. Unlike Fano decoding, the complexity of SCL decoding is independent of the SNR and is of more interest for practical applications. In this paper, we focus on the SCL decoding of PAC codes. Our goal is to reduce the decoding latency of SCL decoding by developing a simplified SCL (SSCL) decoding based on the principles of the SSCL decoding of polar codes. The SSCL decoding of polar codes were proposed and studied in \cite{SSCAlamdar}-\cite{SSCLSarkis2}.
	
%\vspace{-.6cm}
Our contribution in this paper are as follows. We design a SSCL decoder for PAC codes to avoid processing every node in the decoding tree. Instead, at a \emph{special} node, the decoder outputs the decoding result without processing the nodes in the sub-tree of the node. A special node is defined for the information-carrier vector, i.e., CC encoder input, in the precise way it is defined for polar codes. The extension of the special node processing of polar codes to PAC codes is not trivial due to effects of CC encoding. We provide algorithms to process the special nodes in the presence of the CC encoder for four types of special nodes and show how the teh decoding latency can be reduced by up to 34 percent. To mitigate complexity, an \emph{inverse}  CC encoding is introduced to get the list members at the input of the CC encoder. We show that the inverse CC can be implemented with linear complexity in the special node length. We also show that the generator polynomial of the inverse CC has a nested property which allows for lower implementation complexity.
	
The rest of the paper is organized as follows. An overview of PAC code and its SCL decoding algorithm is presented in Section II. In Section III, we present the SSCL decoding algorithm for PAC codes which includes the decoding procedures for four types of special nodes. Numerical results on the performance and complexity are presented in Section IV. Finally, Section V concludes the paper. 
	
\section{Background}
\subsection{Arikan's PAC Codes}
	
A PAC code of length $N=2^n$ with $K$ information bits is denoted by $PAC(N,K, \mathcal{A}, \mathbf{g})$, where $\mathcal{A} \subset \{ 0,\ldots, N-1 \} $ is the information index set with cardinality $K$ and $\mathbf{g}=(g_0, \ldots, g_{m})$ is the generator polynomial of the CC with a \emph{constraint length} of $m+1$. A PAC encoder encodes a binary information vector $\mathbf{d}_0^{K-1}=[d_0, \ldots, d_{K-1}]$ as follows. Let $\mathbf{v}=v_0^{N-1}$ be a binary vector of size $1\times N$ referred to as the information-carrier vector. The $K$ information bits are placed in those elements of $\mathbf{v}$ corresponding to the set $\mathcal{A}$ and zero values are placed in the other $N-K$ elements. This step is called \emph{rate-profiling}. The information-carrier vector in the next step is encoded by the CC to give vector $\mathbf{u}=u_0^{N-1}$ where 
	%\vspace{-.19cm}
\begin{align}
		u_i  = \sum_{j=0}^{\nu} g_j v_{i-j}
		\label{v2u}
\end{align}
where $v_k$ for $k<0$ is set to $0$ by convention. Eq. (\ref{v2u}) can be described as $\mathbf{u}=\mathbf{v}\mathbf{G}_{cc,n}$ where $\mathbf{G}_{cc,n}$ for any $N=2^n$ is an $N\times N$ upper-triangular Toeplitz matrix. 
\begin{equation}
		\hspace{-.6cm} \mathbf{G}_{{cc,n}} =\begin{bmatrix}
			g_{0}  & g_{1} & g_{2}  & \cdots & g_{m} & 0 & \cdots & 0 \\
			0  & g_{0}  & g_{1}  & g_2 & \cdots & g_{m} &   & 0 \\
			0 & 0  & g_0 & g_1  & \ddots & \cdots & g_{m} & \vdots \\
			\vdots & 0 & \ddots & \ddots & \ddots & \ddots &  &  \vdots \\
			\vdots &  & \ddots & \ddots & \ddots & \ddots & 0 & \vdots\\
			\vdots  &  &  & \ddots & 0  & g_0  &  g_1  & g_{2}\\
			\vdots  & && & 0  & 0  &  g_0  & g_{1}\\
			0 & \cdots &  \cdots & \cdots & \cdots & 0 & 0 & g_0  \\
		\end{bmatrix}
		\label{gcc_mat}
\end{equation}
The transformation from $\mathbf{v}$ to $\mathbf{u}$ can also be described via a function $\mathrm{convTrans(.)}$ with initial an initial length-$m$ all-zero state vector $\mathrm{cState}$, adopted from \cite{pacSCLRowshan} and defined in Algorithm \ref{algpacCC1}. The last step of PAC encoding is to encode $\mathbf{u}$ by polar transform and get the codeword $\mathbf{c}=c_0^{N-1}=\mathbf{u}\mathbf{G}_N$ where $\mathbf{G}_N=\mathbf{F}^{\otimes n}$ and $\mathbf{F}$ is the $2\times 2$ binary kernel.

\begin{algorithm}
		\DontPrintSemicolon
		\caption{PAC convolutional encoding}
		\label{algpacCC1}
		\BlankLine
		\Fn{\FLoop{$\mathbf{v}, \mathbf{g}, \mathrm{cState}$}}{
			
			$N$ $\gets$ $\mathrm{Len}(\mathbf{v})$ $\text{      // Length of input vector $\mathbf{v}$ }$
			
			\For{$i \gets 0$ to $N-1$} {
				
				$(u_i,\mathrm{cState})$ $\gets$ $\mathrm{conv1bTrans}(v_i,\mathrm{cState},\mathbf{g})$
			}
			
			$\mathbf{u} \gets [u_0, \ldots, u_{N-1}]$
			
			\KwRet {$(\mathbf{u}, \mathrm{cState})$}
		}
		\Fn{\FLoopb{$v, \mathrm{currState}, \mathbf{g}$}}{
			$u$ $\gets$ $v.g_0$
			
			\For{$i \gets 1$ to $m$} {
				\If{$g_i=1$}{
					
					$u$ $\gets$ $u+\mathrm{currState}[i-1]$
				}
			}
			
			$\mathrm{nextState}$ $\gets$ $[v,\mathrm{currState}[0,\ldots,m-2]]$
			
			\KwRet {$(u,\mathrm{nextState})$}
		}
\end{algorithm}

\subsection{SCL decoding of PAC codes}
SCL decoding of PAC codes were proposed in \cite{pacSCLArman} and \cite{pacSCLRowshan} and is essentially based on the idea of SCL decoding of polar codes. As an improvement on the SC decoding, instead of making a final decision at each information bit, the SCL decoder creates two possibilities for "0" and "1" estimates of the bit. In particular, for a list size of $L$, the SCL decoder doubles the number of paths at each information bit until it reaches $L$ paths. From this point on, at each information bit, each path is expanded to generate two child paths giving a total of $2L$ child paths. A pruning procedure is performed to keep the $L$ most likely paths and discard the other $L$ paths. The path likelihood is determined according a path metric (PM) which is stored for each path. It is noteworthy that, the SCL decoding is in fact $L$ parallel SC decoders that interact with each other at the leaf nodes and update their paths and corresponding metrics. When a path $\mathbf{v}_0^{i-1}$ with corresponding ${\mathbf{u}}_0^{i-1}$ generates a child path ${\mathbf{v}}_0^{i}$ with corresponding  ${\mathbf{u}}_0^{i}$, the child path metric is calculated as follows.
	%\begin{equation}
	%	\mathrm{PM} \left(\widehat{\mathbf{v}}_0^{i} \right)= \begin{cases*}
		%		\mathrm{PM} \left(\widehat{\mathbf{v}}_0^{i} \right)  & if  $(1-2\widehat{u}_i)\lambda^{(\nu)} > 0$  \\
		%		\mathrm{PM} \left(\widehat{\mathbf{v}}_0^{i} \right) -|\lambda^{(\nu)}| & otherwise
		%	\end{cases*} \nonumber
	%\end{equation}
	\begin{equation}
		\mathrm{PM} \left({\mathbf{v}}_0^{i} \right) = \mathrm{PM} \left({\mathbf{v}}_0^{i-1} \right) +  \frac{1-\mathrm{sgn} \left( (1-2{u}_i)\lambda^{(i)} \right)}{2}	|\lambda^{(i)}|
	\end{equation}
where  $\mathrm{sgn}$ is the sign function and $\lambda^{(i)}$ is the LLR calculated by the SC decoder for bit $u_i$ corresponding to  assuming the previous decoded bits are given by ${\mathbf{v}}_0^{i-1}$. The SCL decoder chooses the length-$N$ path with smallest PM as the final decoded word. CRC-aided SCL (CA-SCL) PAC coding scheme is also obtained by appending CRC bits to the information vector $\mathbf{d}_0^{K-1}$ and selecting the path with smallest PM which passes the CRC as the final decoded word.

\section{Simplified Successive Cancellation List Decoding of PAC Codes}
The SC and SCL decoding of PAC codes are described in \cite{pacSCLArman} and \cite{pacSCLRowshan} by recursive calculation of the LLRs of the bit-channels without explicit viewpoint of the constituent codes of the PAC code which is necessary to describe SSC and SSCL decoders. To describe the SSCL decoding algorithm for PAC code, we first describe the SC decoding on the decoding tree from the viewpoint of constituent codes. This allows us to later employ simplified decoders for the constituent codes. 

\subsection{Recursive SC decoding on the decoding tree}
\begin{figure}[!t]
	\begin{center}
		% Onecolun: \includegraphics[scale=.7,viewport=0 0 270 450,clip]{G_Tanner_Graph.pdf}
		%\includegraphics[scale=.52,viewport=0 15 580 180,clip]{AE_general.PNG}
		%\includegraphics[scale=.42,viewport=15 -12 570 390,clip]{pc8_withcState_final.pdf}
		\includegraphics[scale=.41,viewport=15 -12 570 390,clip]{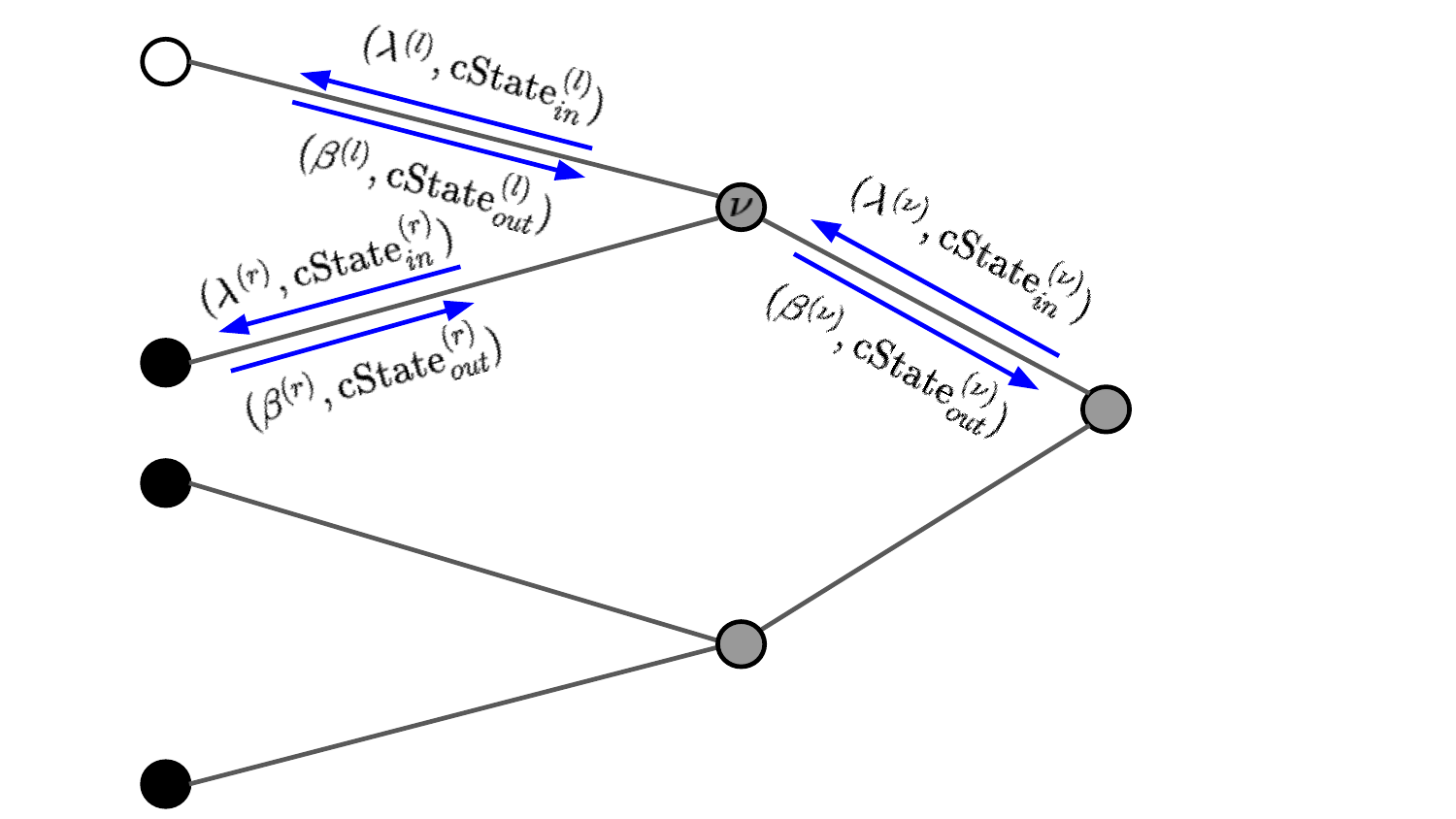}
	\end{center}
	\vspace*{-4mm}
	\caption{Decoding tree of a length-4 PAC code}
	\label{n4dectree}
\end{figure}

The SC Decoding of PAC code can be described on a decoding tree that is traversed depth first, i.e., right to left, where each sub-tree corresponds to a constituent code. The decoding tree of a length-$4$ PAC code with information set $\mathcal{A}=\{1,2,3\}$ is shown in Fig. \ref{n4dectree}. The frozen and information bits are shown as white and black leaf nodes, respectively. The leaf nodes from top to bottom correspond to bits $v_i$ and $u_i$ for $i=0,\ldots,3$. The SC decoder traverses this tree. At a node $\nu$ corresponding to a constituent code of length $N_\nu$ an length-$N_\nu$ input LLR vector  $\mathbf{\lambda}^{(\nu)}=[\lambda_0^{(\nu)}, \ldots,\lambda_{N_{\nu}-1}^{(\nu)} ]$ and a current state vector $\mathrm{cState}_{in}^{(\nu)}$ are used to calculate the decoded codeword $\mathbf{\beta ^ {(\nu)}}=[\beta_0^{(\nu)}, \ldots, \beta_{N_{\nu}-1}^{(\nu)}]$ and a next state vector $\mathrm{cState}_{out}^{(\nu)}$. The state vector is the counterpart of the state vector at the encoder and is to consider the effect of CC encoding at the decoder by storing the state of the CC encoder for the decoded information-carrier vector up to certain decoded bit index. The decoding tree is comprised of $n+1$ columns. From right to left, the columns are indexed from $0$ to $n$. There are $2^j$ nodes at column $j$ each node corresponding to a constituent code of length $2^{n-j}$. The SC decoding of PAC code is performed as follows. The input LLR vector and the current state vector at the node at column $0$ are set to the length-$N$ channel LLR vector and all-zero vector, respectively. When a node $\nu$ receives the length-$N_\nu$ LLR vector $\mathbf{\lambda}^{(\nu)}$, the input LLR and state vector for its left child is calculated as follow.
\begin{align}
		&{\lambda}_i^{(l)} = {\lambda}_i^{(\nu)} \boxplus {\lambda}_{i+N_{\nu}/2}^{(\nu)}, \text{ for } i=0,\ldots, N_{\nu}/2 \nonumber \\
		&\mathrm{cState}_{in}^{(l)} =\mathrm{cState}_{in}^{(\nu)}
\end{align}
The decoding of the left child node is performed and the output codeword and state are obtained as $\mathbf{\beta}^{(l)}$ and $\mathrm{cState}_{out}^{(l)}$, respectively. The input LLR and state vector for the right child node are then calculated as 
\begin{align}
		&{\lambda}_i^{(r)} = (1-2\beta_i^{(l)}){\lambda}_i^{(\nu)} + {\lambda}_{i+N_{\nu}/2}^{(\nu)} \text{ for } i=0,\ldots, N_{\nu}/2 \nonumber \\
		&\mathrm{cState}_{in}^{(r)} = \mathrm{cState}_{out}^{(l)}
\end{align}
Once the outputs $\mathbf{\beta}^{(r)}$ and $\mathrm{cState}_{out}^{(r)}$ of the right child node are available, the outputs of node $\nu$ are calculated as
	\begin{equation}
		\beta_{i}^{(\nu)} = \begin{cases*}
			\beta_{i}^{(l)} + \beta_{i}^{(r)} & if  $i<N_{\nu}/2$  \\
			\beta_{i-N_{\nu}/2}^{(r)} & if $i  \geq N_{\nu}/2$
		\end{cases*} \nonumber
	\end{equation}
	\begin{equation}
		\hspace{-1.85cm} \mathrm{cState}_{out}^{(\nu)} =\mathrm{cState}_{out}^{(r)}
	\end{equation}

At a leaf node corresponding to an formation index $[z_j, \mathrm{nextState}_j]=\mathrm{conv1bTrans}(j, \mathrm{cState}_{in}^{(\nu)},\mathbf{g})$ is calculate for $j \in \{0,1\}$. Then, $\mathbf{\beta}^{(\nu)}$ and $\mathrm{cState}_{out}^{(\nu)}$ are calculated as
	\begin{align}
		&\mathbf{\beta}^{(\nu)} = z_{j^\star}  \nonumber \\
		& \mathrm{cState}_{out}^{(\nu)} = \mathrm{nextState}_{j^\star} 
		\label{dec_leafe}
	\end{align}
	where $j^\star = \mathrm{argmax}_j (1-2z_j)\lambda^{(\nu)}$. At a frozen node, $\mathbf{\beta}^{(\nu)}$ and $\mathrm{cState}_{out}^{(\nu)}$ are calculated according to Eq. (\ref{dec_leafe}) with $j^\star =0$.
	
	The decoded information-carrier vector $\widehat{\mathbf{v}}=\widehat{v}_0^{N-1}$ and the corresponding vector $\widehat{\mathbf{u}}=\widehat{u}_0^{N-1}$ are obtained according to the decoding results at leaf nodes. In particular, at a leaf node $\nu$ corresponding to index $k$, we set $\widehat{v}_k = j^{\star}$ and $\widehat{u}_k = z_{j^{\star}}$.

The idea behind the SSCL decoder of PAC codes is similar to that of polar codes. The SSCL decoder traverses decoding tree as in the same way as SCL decoding while performing operations described in this section and immediately moving to their parent nodes when it encounters a special node $\nu$. Compared to polar codes, the processing of a special node for PAC code needs consideration of the effects of CC encoding. Compared to Fig. \ref{n4dectree}, we drop the notation dependency on $\nu$ for simplicity. At each node it is assumed that current state $\mathrm{cState}_{in}$, input LLR vector $\mathbf{\lambda}$ and the corresponding metric $\mathrm{PM}$ of the parent path are available for each member in the current list. We are interested in determining $Z$ length-$N_{\nu}$, $N_{\nu}=2^{n_{\nu}}$, candidates for node $\nu$ containing the bits of the information-carrier vector corresponding to the leaf nodes of $\nu$. For simplicity we define these vectors starting at index $0$, e.g., $\mathbf{v}^{(z)} = [v_0^{(z)}, \ldots, v_{N_{\nu}-1}^{(z)}]$, for $z=0,\ldots,Z-1$. Let's denote the output of the CC encoder, the decoded codeword of the constituent code at the node and the output state by $\mathbf{u}^{(z)} = [u_0^{(z)}, \ldots, u_{N_{\nu}-1}^{(z)}]$, $\mathbf{\beta}^{(z)}=[\beta_0^{(z)}, \ldots, \beta_{N_{\nu}-1}^{(z)}]$ and $\mathrm{cState}_{out}^{(z)}$, respectively. The goal is to determine $\mathbf{v}^{(z)}$, $\mathbf{u}^{(z)}$, $\mathbf{\beta}^{(z)}$ and $\mathrm{cState}_{out}^{(z)}$ for the node. Once $\mathbf{\beta}^{(z)}$ is determined, the PM for the candidate $z$ is calculated as 
\begin{equation}
	\mathrm{PM}^{(z)} = \mathrm{PM}+ \sum_{j=0}^{N_{\nu}-1} \frac{1-\mathrm{sgn} \left( ( 1-2\beta_{j}^{(z)})\lambda_j^{(\nu)} \right)}{2} |\lambda_{j}^{(\nu)}|	
\end{equation}

Once $Z$ candidates are generated for each member in the list, the expanded list is pruned to up to $L$ list member with smallest PM, if it contains more than $L$ members. In the following we consider four types of special nodes.	
	
\subsection{Rate-0 node }
A rate-0 node is a node for which all the leaf nodes are frozen. One candidate, i.e., $Z=1$, is generated as $\mathbf{v}^{(0)}=\mathbf{0}$. Unlike the polar code the output $\mathbf{\beta}^{(0)}$ of the node is not necessarily all-zero due to the CC code. To obtain the output, we first calculate $(\mathbf{u}^{(0)}, \mathrm{cState}_{out}^{(0)}) = \mathrm{convTrans} \left( \mathbf{0}, \mathbf{g}, \mathrm{cState}_{in} \right)$, where the function $\mathrm{convTrans}$ is defined in Algorithm~\ref{algpacCC1}. The output of the node is calculated as $\mathbf{\beta}^{(0)}=\mathbf{u}^{(0)}\mathbf{F}^{\otimes n_{\nu}}$.

	\subsection{Repetition node}
	
	A repetition node of length $N_{\nu}$ is a node for which the first $N_{\nu}-1$ leaf nodes are frozen and the last one is information. Two child paths are generated corresponding to the two possibilities for the last bits. That is, $\mathbf{v}^{(0)}=[\mathbf{0}_{1 \times N_{\nu}-1}, 0]$ and $\mathbf{v}^{(1)}=[\mathbf{0}_{1 \times N_{\nu}-1}, 1]$ where $\mathbf{0}$ is an  all-zero vector of length $N_{\nu}-1$. For the first child, $(\mathbf{u}^{(0)}, \mathrm{cState}_{out}^{(0)}) = \mathrm{convTrans} \left( \mathbf{0}, \mathbf{g}, \mathrm{cState}_{in} \right)$ and $\mathbf{\beta}^{(0)}=\mathbf{u}^{(0)}\mathbf{F}^{\otimes n_{\nu}}$ as calculated for the Rate-0 node. For the second child, decoding output can be directly calculated as follows, without calling $\mathrm{convTran(.)}$ with $\mathbf{v}^{(1)}$ as the input.
	\begin{align}
		&\mathbf{u}^{(1)} = \mathbf{u}^{(0)} + [\mathbf{0}_{1\times N_{\nu}-1}, 1] \nonumber \\
		&\mathrm{cState}_{out}^{(1)} = \mathrm{cState}_{out}^{(0)} + [1, \mathbf{0}_{1\times m-1}] \nonumber \\
		&\mathbf{\beta}^{(1)} = \mathbf{\beta}^{(0)} + \mathbf{1}_{1 \times N_{\nu}}
		\label{repEq}
	\end{align}
	
\subsection{Rate-1 node}
A rate-1 node is a node for which all the leaf nodes are information nodes. In the presence of the CC encoder, the constituent code corresponding to the node is still  rate-1 code. To see why, since the CC is a linear operator we can write
\begin{equation}
		\mathbf{u}^{(z)} = \mathbf{v}^{(z)} \mathbf{G}_{cc,n_{\nu}}+ \mathbf{\eta} \label{rate1eq}
\end{equation}  
where $\mathbf{G}_{cc, n_\nu}$ is the $N_{\nu} \times N_{\nu}$ matrix given in Eq. (\ref{gcc_mat}) and $\mathbf{\eta}$ is a the output vector of CC encoder with all-zero input and an initial state given by $\mathrm{cState}_{in}$. It is straightforward to see that $\mathbf{u}^{(z)}$ can take any value in $\mathrm{GF}(2)^{N_{\nu}}$ and so does $\mathbf{\beta}^{(z)}=\mathbf{u}^{(z)} \mathbf{F}^{\otimes n_{\nu}}$. Therefore, the routines for generating candidates of a rate-1 nodes in case of polar SSCL decoding can be reused, see [12, Sec. III.C]. Let's denotes such a routine by $\mathrm{getRate1Candidate(.)}$. This routine takes a LLR vector $\mathbf{\lambda}$ and outputs $Z$ most likely codewords. The $\mathbf{\beta}^{(z)}$ and $\mathbf{u}^{(z)}$ are determined as 
	\begin{align}
		&[\mathbf{\beta}^{(0)}, \ldots, \mathbf{\beta}^{(Z-1)}] = \mathrm{getRate1Candidate} \left( \mathbf{\lambda} \right)  \\
		& \mathbf{u}^{(z)} = \mathbf{\beta}^{(z)}\mathbf{F}^{\otimes n_{\nu}} \label{beta2v}
	\end{align}
where (\ref{beta2v}) holds because polar transform matrix is involutory. To get $\mathbf{v}^{(z)}$, (\ref{rate1eq}) can be solved for given $\mathbf{v}^{(z)}$ and $\mathbf{\eta}$, i.e., $\mathbf{v}^{(z)}= (\mathbf{u}^{(z)}+\mathbf{\eta}) \mathbf{G}_{cc,n_\nu}^{-1}$. 	
In Lemma \ref{mylemma}, we show that $\mathbf{G}_{cc,n_\nu}^{-1}$ is also an upper-triangular Toeplitz matrix and has a nested property with respect to the special node length $N_\nu$.

% with an \emph{inverse generator polynomial} $\mathbf{g}^{-1}$. 

\begin{lemma}
For any upper-triangular Toeplitz matrix $\mathbf{G}_{cc,n_\nu}$, the inverse matrix $\mathbf{G}_{cc,n_\nu}^{-1}$ is an upper-triangular Toeplitz matrix and takes the following form
\begin{align}
	&\mathbf{G}_{cc,n_\nu+1}^{-1} =\begin{bmatrix}
		\mathbf{G}_{cc,n_\nu}^{-1}  & \mathbf{P}_{n_\nu}  \\
		0  & \mathbf{G}_{cc,n_\nu}^{-1}  \\
	\end{bmatrix}
	\label{fsdfsd} 
\end{align}
fro some $N_\nu \times N_\nu$ matrix $\mathbf{P}_{n_\nu}$.
\label{mylemma}
\end{lemma}
We omit the proof. Algorithm \ref{algrate1} summarizes the decoding of rate-1 node. The parameter $q$ is defined as $q=\mathrm{min}(N_{\nu},m)$.

\begin{algorithm}
		\SetKwInOut{Input}{input}\SetKwInOut{Output}{output}
		\Input{LLR vector $\mathbf{\lambda}$, input state $\mathrm{cState}_{in}$, $\mathbf{G}_{cc,n_\nu}^{-1}$}
		\Output{$\mathbf{v}^{(z)}$, $\mathbf{u}^{(z)}$, $\mathbf{\beta}^{(z)}$, $\mathrm{cState}_{out}^{(z)}$, $z=0,\ldots, Z-1$}
		\DontPrintSemicolon
		\caption{Rate-1 node decoding of PAC code}
		\label{algrate1}
		
		$[\mathbf{\beta}^{(0)}, \ldots, \mathbf{\beta}^{(Z-1)}]$ $\gets$ $\mathrm{getRate1Candidates} \left( \mathbf{\lambda} \right)$

		$\left(\mathbf{\eta}, \sim \right)$ $\gets$ $\mathrm{convTrans} \left( \mathbf{0}, \mathbf{g}, \mathrm{cState}_{in} \right)$
		
		\For{$z \gets 0$ to $Z-1$} {
			
			$\mathbf{u}^{(z)}$ $\gets$ {$\mathbf{\beta}^{(z)}\mathbf{F}^{\otimes n_{\nu}}$}
			
			$\mathbf{v}^{(z)}$ $\gets$ {$(\mathbf{u}^{(z)}+\mathbf{\eta}) \mathbf{G}_{cc,n_\nu}^{-1}$}

			$\mathrm{cState}_{out}^{(z)}$ $\gets$ $[v_{N_{\nu}-1}^{(z)},\ldots,v_{N_\nu-q}^{(z)},\mathrm{currState}[0,\ldots,m-q-1]]$
		}
\end{algorithm}

\subsection{Single Parity Check (SPC) node}
A SPC node of length $N_{\nu}$ is a node for which the first leaf node is frozen and the remaining $N_{\nu}-1$ leaf nodes are information. We show that in the presence of the CC encoder, the output $\mathbf{\beta}^{(z)}$ of a node no longer takes value from the codebook of a SPC code.  Instead, the output takes value from a SPC codebook plus a constant vector. To see why, from Eq. (\ref{rate1eq}) the output of the node is
	\begin{align}
		&\mathbf{\beta}^{(z)}=\mathbf{u}^{(z)}\mathbf{F}^{\otimes n_{\nu}} =\mathbf{v}^{(z)} \mathbf{G}_{cc,n_\nu}.\mathbf{F}^{\otimes n_{\nu}}+ \mathbf{\eta}\mathbf{F}^{\otimes n_{\nu}}
	\end{align}
The first element of $\mathbf{v}^{(z)} \mathbf{G}_{cc,n_\nu}$ is always 0 due to $v_0^{(z)}=0$ corresponding to a frozen leaf node. Therefore $\mathbf{v}^{(z)} \mathbf{G}_{cc,n_\nu}\mathbf{F}^{\otimes n_{\nu}}$ takes value from a SPC codebook. Defining the coded $\mathbf{\eta}$ as $\mathbf{\eta}_{c}=[\eta_{c,0}, \ldots, \eta_{c,N_{\nu}-1}]=\mathbf{\eta}\mathbf{F}^{\otimes n_{\nu}}$, we can remove the impact of $\mathbf{\eta}_{c}$ from the LLR vector and still benefit from decoding techniques available for a SPC code. The impact of $\mathbf{\eta}_{c}$ can be removed by multiplying the LLR values $\lambda_{j}^{(\nu)}$ with $\pm 1$ depending on the value of $\eta_{c,j}$. Once the effects of $\mathbf{\eta}_c$ is removed, the LLR vector can be passed to a routine which generates the candidates of SPC code, see [12, Sec. III.D]. Let's denote such a routine by $\mathrm{getSPC\_Candidate(.)}$. This routine takes a LLR vector $\mathbf{\lambda}$ and outputs $Z$ most likely SPC codewords. Adding $\mathbf{\eta}_c$ to these codewords gives the $Z$ candidate codewords at the output of the node. Algorithm \ref{spcalg} summarizes the decoding of a SPC node.

\begin{algorithm}
		\SetKwInOut{Input}{input}\SetKwInOut{Output}{output}
		\Input{LLR vector $\mathbf{\lambda}$, input state $\mathrm{cState}_{in}$, $\mathbf{G}_{cc,n_\nu}^{-1}$}
		\Output{$\mathbf{v}^{(z)}$, $\mathbf{u}^{(z)}$, $\mathbf{\beta}^{(z)}$, $\mathrm{cState}_{out}^{(z)}$, $z=0,\ldots, Z-1$}
		\DontPrintSemicolon
		\caption{SPC node decoding of PAC code}
		\label{spcalg}

		%\State {\mathrm{cState}_{in,copy}} $\gets$ \mathrm{cState}_{in}
		
		$(\mathbf{\eta}, \sim)$  $\gets$ $\mathrm{convTrans} \left( \mathbf{0}, \mathbf{g}, \mathrm{cState}_{in} \right)$
		
		$\mathbf{\eta}_{c}$ $\gets$ $\mathbf{\eta}\mathbf{F}^{\otimes n_{\nu}}$ 	$  \text{      and  } \tilde{\mathbf{\lambda}}$ $\gets$ $\left(1-2\mathbf{\eta}_{c}\right) \mathbf{\lambda}$
		
		$[\tilde{\mathbf{\beta}}^{(0)}, \ldots, \tilde{\mathbf{\beta}}^{(Z-1)}]$ $\gets$ $\mathrm{getSPC\_Candidate} ( \tilde{\mathbf{\lambda}})$
		
		\For{$z \gets 0$ to $Z-1$} {
			
			$\mathbf{\beta}^{(z)}$ $\gets$ $\tilde{\mathbf{\beta}}^{(z)} + \mathbf{\eta}_{c}$

			$\mathbf{u}^{(z)}$ $\gets$ {$\mathbf{\beta}^{(z)}\mathbf{F}^{\otimes n_{\nu}}$}
			
			%$(\mathbf{v}^{(z)}, \sim)$ $\gets$ {$\mathrm{convTrans}(\mathbf{u}^{(z)} %+\mathbf{\eta}, \mathbf{g}^{-1})$}
			
			$\mathbf{v}^{(z)}$ $\gets$ {$(\mathbf{u}^{(z)}+\mathbf{\eta}) \mathbf{G}_{cc,n_\nu}^{-1}$}

			$\mathrm{cState}_{out}^{(z)}$ $\gets$ $[v_{N_{\nu}-1}^{(z)},\ldots,v_{N_\nu-q}^{(z)},\mathrm{currState}[0,\ldots,m-q-1]]$
		}
\end{algorithm}

In line 2, $\left(1-2\mathbf{\eta}_{c}\right) \mathbf{\lambda}$ is obtained by component-wise product of vector $1-2\mathbf{\eta}_{c}$ and $\mathbf{\lambda}$. 
	
\section{Numerical Results }

\subsection{Performance}
\begin{figure}[!t]
		\begin{center}
			\includegraphics[scale=.58,viewport=100 212 865 568,clip]{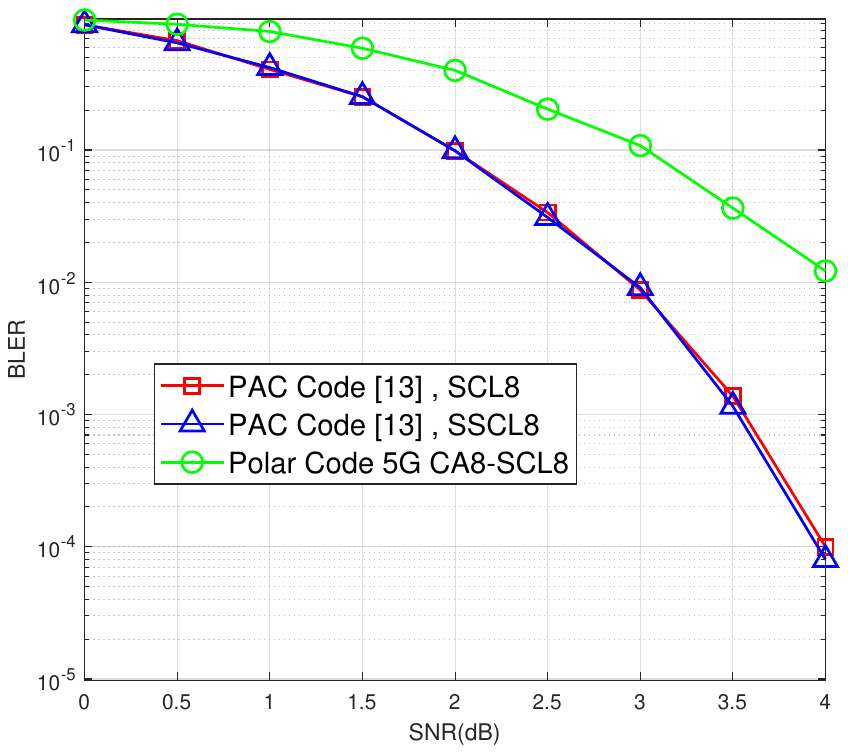}
		\end{center}
		%\vspace*{-5mm}
		\caption{BLER performance of (128,72) PAC code and CRC-aided polar code under SCL decoding.}
		\label{n128k72pacscl8}
	\end{figure}

	We begin with an investigation in to the performance of PAC codes under SSCL decoding. First, we consider a $(128, 72)$ PAC codes with generator polynomial $\mathbf{g}=(1,0,1,1,0,1,1)$. The rate profile is adopted from \cite{pacQRL_SKMishra} and the transmission is over BI-AWGN channel with BPSK signaling and noise variance $\sigma^2$. SNR is defined as $\mathrm{SNR}=10\mathrm{log}_{10} 1/\sigma^2$. For SSCL decoder we set $Z=4$. Fig. \ref{n128k72pacscl8} depicts the BLER performance of the PAC code as well as the 5G 3GPP NR polar code with 8 bit CRC \cite{ts38212}. As can be seen, the performance of PAC code under SCL and SSCL decoding is almost identical with SSCL showing minute gain over SCL. Due to extremely close performance, we have set the same noise realizations for both decoders for more reliable estimation of the BLER.

\begin{figure}[!t]
		\begin{center}
			\includegraphics[scale=.58,viewport=100 212 865 568,clip]{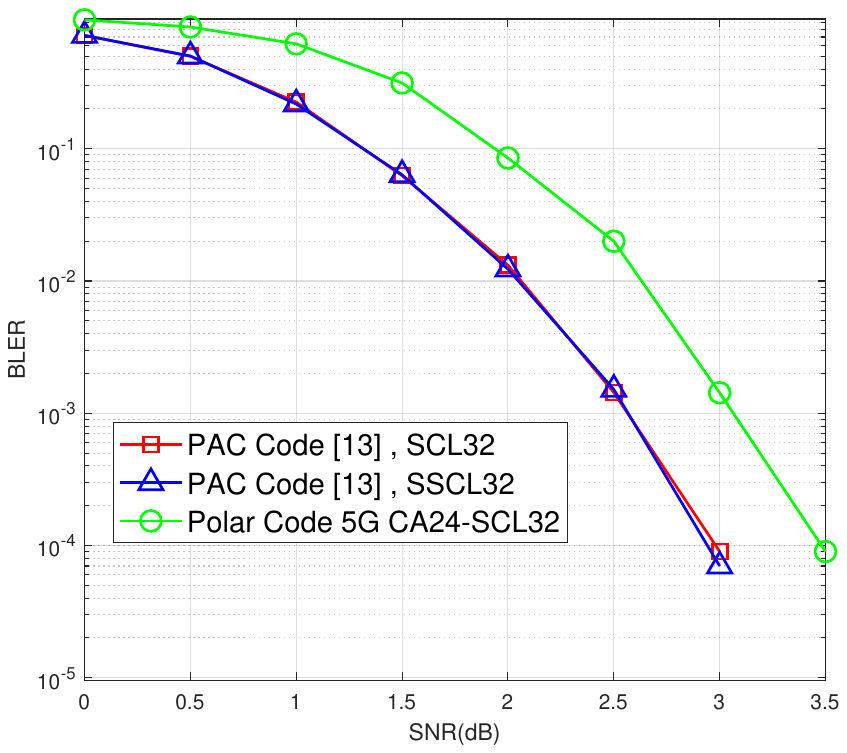}
		\end{center}
		%\vspace*{-5mm}
		\caption{BLER performance of (256,128) PAC code and CRC-aided polar code under SCL decoding.}
		\label{n256k128pacscl32}
	\end{figure}

	We also consider a larger $(256,128)$ PAC code with generator polynomial $\mathbf{g}=(1,0,1,1,0,1,1)$. The rate profile is adopted from \cite{pacQRL_SKMishra} and is optimized for SCL decoding with a list size of 32. Fig. \ref{n256k128pacscl32} shows the resulting BLER performance. It can be seen that the similar performance between the SCL and SSCL decoding can also be observed for this code. Similar to the previous code the noise realization is the same for both codes and negligible performance difference is observed between the two decoders. 
	
\subsection{Complexity}

The decoding latency analysis is based on the following assumption. In conventional SCL decoding it takes $2N-2$ cycles to calculates the LLRs of every node in the graph \cite{SSCLHashemi1}. Path metric update, sorting and pruning are assumed to take one cycle. Partial encoding, i.e., generating the codeword of parent node from those of the child nodes, is assumed to be done instantaneously. This leads to a total of $2N-2+K$ cycles in the case of polar code. The number of cycles for PAC SCL decoding is also $2N-2+K$. To see why, we note that once the LLR of $u_i$ corresponding to a leaf node is calculated, list expansion, pruning, updating $\mathrm{cState}_{out}$ and partial encoding can be carried out without the need of waiting for calculating the corresponding $v_i$.
%, hence the number of cycles is the same as PAC SCL.

%Since it takes at least one cycle to calculate the LLR of next leaf %node, $v_i$ will be ready when decoding reaches the next leaf node.

\begin{table}[t]
	\centering
	\caption{Time step complexity comparison for the PAC codes }
	\begin{tabular}{|c|c|c|c|}
		\hline
		PAC Code & \shortstack{Time Steps \\ (SCL)} & \shortstack{Time Steps \\ (SSCL)} & \shortstack{Reduction \\ percentage}\\
		\hline
		\shortstack{ PAC(128,72) \cite{pacQRL_SKMishra} } & 321 & 217 & 32.3\\
		\hline
		\shortstack{ PAC(256,128) \cite{pacQRL_SKMishra} } & 633 & 373 & 41.0\\
		\hline
	\end{tabular}
	\label{jadval}
\end{table}
With PAC SSCL decoding, operating at the special node level saves the number of cycles required for calculating the LLRs of the subsequent nodes. For Rate-0 node $\mathrm{convTrans()}$ needs to run to get the output state which only requires $q=\mathrm{min}(N_{\nu},m)$ cycles due to all-zero input if $N_{\nu}>m$. This also holds for Repetition node due to use of (\ref{repEq}). Since PM sorting is needed for repetition node, the total number of cycles for Rate-0 and repetition node are $q$ and $q+1$, respectively. For Rate-1 and SPC node, candidate generation and PM sorting is assumed to take $\kappa+1$ cycles. $\mathbf{u}^{(z)}$ is assumed to be done instantaneously. calculation of $\mathbf{\eta}$ can be done in $q$ cycles. Together, candidate generation, PM sorting and calculation of $\mathbf{\eta}$ can be done in $\mathrm{max}\left(\kappa+1,q\right)$ cycles as the calculation of $\mathbf{\eta}$ can be done in parallel. This gives a total of $\mathrm{max}\left(\kappa+1,q\right)$ cycles for Rate-1 node and is also the number of cycles needed for decoding the SPC node. Table \ref{jadval} shows the required time steps in cycles for the considered PAC codes assuming $\kappa=1$. As can be seen, the SSCL decoding can save up to 41 percentage of cycles. The corresponding statistics for the special nodes in the decoding tree is shown in Tables \ref{jadval3} and \ref{jadval4}.

\begin{table}[t]
	\centering
	\caption{Special node statistics for PAC(128,72)}
	\begin{tabular}{|c|c|c|c|c|}
		\hline
		Node length & Rate-0 & Repetition  & Rate-1  & SPC \\
		\hline
		2 & 1 & 2 & 1 & 0\\
		\hline
		4 & 0 & 7 & 1 & 6\\
		\hline
		8 & 1 & 2 & 1 & 2\\
		\hline
		16 & 0 & 0 & 0 & 1\\
		\hline
	\end{tabular}
	\label{jadval3}
\end{table}

\begin{table}[t]
	\centering
	\caption{Special node statistics for PAC(256,128)}
	\begin{tabular}{|c|c|c|c|c|}
		\hline
		Node length & Rate-0 & Repetition  & Rate-1  & SPC \\
		\hline
		2 & 3 & 14 & 3 & 0\\
		\hline
		4 & 3 & 5 & 5 & 3\\
		\hline
		8 & 0 & 2 & 2 & 3\\
		\hline
		16 & 0 & 4 & 0 & 0\\
		\hline
		32 & 0 & 0 & 0 & 1\\
		\hline
	\end{tabular}
	\label{jadval4}
\end{table}

Compared to the SCL decoding, the main introduced overhead is the multiplication with $\mathbf{G}_{cc,n_\nu}^{-1}$ for the rate-1 and SPC node. The computational complexity of this operation cannot be reduced unless the generator polynomial has a short length which is not guaranteed even for small values of $m$. However, the nested property allows us to only employ one multiplication circuit for the largest special node length of interest $N_{\nu,max}$ and employ it for any length $N_{\nu}\leq N_{\nu, max}$ which can reduce the area complexity of the decoder.

\section{Conclusion}
In this paper we proposed an SSCL decoder for PAC codes based on the underlying principles of SSCL decoder for polar codes. The proposed SSCL decoder applies includes operations to acknowledge the effects of CC encoding in the encoder. We provided decoding details for four types of special nodes. We also provided an analysis on the decoding latency of the SSCL decoder and showed that the latency can be reduced by almost 41\% without meaningful impact on the BLER performance.

	\section{Appendix}
	\subsection{Proof of Lemma \ref{mylemma}}
	
	In Lemma \ref{mylemma}, we show that the inverse of an full-rank $N \times N$, upper-triangular Toeplitz matrix $\mathbf{G}_{N}$ with the generator polynomial $\mathbf{g}=(g_0, \ldots, g_m)$, is also an upper-triangular Toeplitz matrix with an \emph{inverse generator polynomial} $\mathbf{g}^{-1}=(\alpha_0, \ldots, \alpha_{N-1} )$. 
	
	We prove the lemma by induction. Note that $g_0=g_m=\alpha_0=1$ as $\mathbf{G}_N$ and $\mathbf{G}_N^{-1}$ are full-rank. The result holds for $N=2$ with $\mathbf{G}_N^{-1}$ given as below.
	
	\begin{equation}
		\mathbf{G}_{2}^{-1} =\begin{bmatrix}
			\alpha_0  & \alpha_1  \\
			0  & \alpha_0   \\
		\end{bmatrix}
		\label{gcc_mat3}
	\end{equation}

	with $\alpha_0 =1$ and $\alpha_1=g_1$. Next, we assume that the result holds for $N$ and prove that it also holds for $N+1$. We have
	
	\begin{equation}
		\mathbf{G}_{N+1} =\begin{bmatrix}
			\mathbf{G}_{N}  & \begin{bmatrix}
				\mathbf{0}  \\
				g_m  \\
				\vdots \\
				g_1
			\end{bmatrix}  \\
			0  & g_{0}  \\
		\end{bmatrix}
		\label{gcc_mat3}
	\end{equation}
	The inverse matrix can be written in the following form. 
	\begin{equation}
		\mathbf{G}_{N+1}^{-1} =\begin{bmatrix}
			\mathbf{B}_{N}  & \mathbf{c} \\
			\mathbf{0}  & d  \\
		\end{bmatrix}
		\label{gcc_mat2}
	\end{equation}
	Since $\mathbf{G}_{N+1}^{-1}.\mathbf{G}_{N+1}=\mathbf{I}_{N+1}$, the following three equations are obtained.
	\begin{align}
		& \mathbf{B}_N \mathbf{G}_N =\mathbf{I}_N \rightarrow \mathbf{B}_N = \mathbf{G}_N^{-1} \label{ap1} \\
		& \mathbf{B}_N. \begin{bmatrix}
			\mathbf{0}  \\
			g_m  \\
			\vdots \\
			g_1
		\end{bmatrix} +\mathbf{c} =\mathbf{0}_{N\times 1} \rightarrow \mathbf{c} =\mathbf{G}_N^{-1}.\begin{bmatrix}
			\mathbf{0}  \\
			g_m  \\
			\vdots \\
			g_1
		\end{bmatrix} \label{ap2} \\
		& dg_0 =1 \rightarrow d=1=\alpha_0 
	\end{align}
	Therefore, the Toeplitz structure of $\mathbf{G}_{N+1}^{-1}$ appears in $\mathbf{B}_N$. That is, the first $N$ elements of the first row of $\mathbf{G}_{N+1}^{-1}$ are $[\alpha_0, \ldots, \alpha_{N-1}]$. To complete the proof, it suffices to show $\mathbf{c}= [c_1, \ldots, c_N]^T= [\alpha_N, \ldots, \alpha_1]^T$ for some $\alpha_N$. In order to prove this, we note that from $\mathbf{G}_N^{-1}.\mathbf{G}_N=\mathbf{I}_N$, the first $N-1$ rows of $\mathbf{G}_N^{-1}$ are orthogonal to the last column of $\mathbf{G}_N$. That is
	\begin{equation}
		[\mathbf{0}, \alpha_0, \ldots, \alpha_{N-j+1}]\begin{bmatrix}
			\mathbf{0}  \\
			g_m  \\
			\vdots \\
			g_1 \\
			g_0
		\end{bmatrix} =0 , \text{      for  }2\leq j \leq N \label{ole231}
	\end{equation} 
	From Eq. (\ref{ap2}) we have $c_j=[\mathbf{0}, \alpha_0, \ldots, \alpha_{N-j}].[\mathbf{0}, g_m, \ldots, g_1]^T$, it follows from (\ref{ole231}) that  $c_j+\alpha_{N-j+1}.g_0 =0$, or
	\begin{equation}
		c_j =\alpha_{N-j+1} \text{ for } 2 \leq j \leq N
	\end{equation}
	The value of $\alpha_N$ is
	\begin{equation}
		\alpha_N  = c_1 =[ \alpha_0, \ldots, \alpha_{N-1}]\begin{bmatrix}
			\mathbf{0}  \\
			g_m  \\
			\vdots \\
			g_1 
		\end{bmatrix}
	\end{equation}
	which completest the proof that $\mathbf{G}_{N+1}^{-1}$ is upper-triangular Toeplitz with generator polynomial $(\alpha_0, \ldots, \alpha_{N-1}, \alpha_N)$. Therefore, the statement holds for all values of $N$.

	We now prove that the inverse matrices are nested for different length of special nodes. To prove this, we can write $\mathbf{G}_{cc,m+1}$ and $\mathbf{G}_{cc,m+1}^{-1}$ as
	\begin{align}
		&\mathbf{G}_{cc,m+1} =\begin{bmatrix}
			\mathbf{G}_{cc,m}  & \mathbf{A}  \\
			0  & \mathbf{G}_{cc,m}  \\
		\end{bmatrix}
		\label{fsdfsd} \\
		& \mathbf{G}_{cc,m+1}^{-1} =\begin{bmatrix}
			\mathbf{B}  & \mathbf{P}  \\
			0  & \mathbf{D}  \\
		\end{bmatrix}.
	\end{align}
	Multiplying $\mathbf{G}_{cc,m+1}^{-1}$ and $\mathbf{G}_{cc,m+1}$ to get the identity matrix $\mathbf{I}_{m+1}$ of size $2^{m+1} \times 2^{m+1}$ results in 
	\begin{equation}
		\mathbf{B}.\mathbf{G}_{cc,m}= \mathbf{I}_m \rightarrow \mathbf{B} = \mathbf{G}_{cc,m}^{-1}
	\end{equation}
Multiplying the second row of $\mathbf{G}_{cc,m+1}^{-1}$ with the second column of $\mathbf{G}_{cc,m+1}$ gives

\begin{equation}
	\mathbf{D}.\mathbf{G}_{cc,m}= \mathbf{I}_m \rightarrow \mathbf{D} = \mathbf{G}_{cc,m}^{-1}
\end{equation}
Finally, multiplying the first row with the second column gives
\begin{equation}
	\mathbf{B}.\mathbf{A} + \mathbf{P}\mathbf{G}_{cc,m} = \mathbf{0} \rightarrow \mathbf{P} = \mathbf{G}_{cc,m}^{-1}.\mathbf{A}.\mathbf{G}_{cc,m}^{-1}.
\end{equation}
That is

\begin{align}
	&\mathbf{G}_{cc,m+1}^{-1} =\begin{bmatrix}
		\mathbf{G}_{cc,m}^{-1}  & \mathbf{G}_{cc,m}^{-1}.\mathbf{A}.\mathbf{G}_{cc,m}^{-1}  \\
		0  & \mathbf{G}_{cc,m}^{-1} \\
	\end{bmatrix}
\label{chisebara}
\end{align}
This completes the proof of Lemma \ref{mylemma}.

\begin{table}
	\centering
	\caption{Nested property of inverse generator polynomial $  \text{                           for  } \mathbf{g}=(1,0,1,1,0,1,1)$}
	\begin{tabular}{|c|c|}
		\hline
		Special node length  &  inverse generator polynomial\\
		\hline
		2 &  (1,0)\\
		\hline
		4 &  (1,0,1,1)\\
		\hline
		8 &  (1,0,1,1,1,1,1,1)\\
		\hline
		16 &  (1,0,1,1,1,1,1,1,0,0,1,0,1,0,1,0)\\
		\hline
	\end{tabular}
	\label{jadval2}
\end{table}

From (\ref{chisebara}), it can also be seen that the inverse generator polynomial for a length- $2^{m}$ node is embedded in the inverse generator polynomial of a length-$2^{m+1}$ node. That is, the first $2^m$ elements of the inverse generator polynomial of a length -$2^{m+1}$ node are the inverse generator polynomial of a length-$2^m$ node. The inverse generator polynomials for the generator polynomials $\mathbf{g}=(1,0,1,1,0,1,1)$ for different node lengths are shown in Table \ref{jadval2}.

\end{document}